\documentclass[sn-basic]{sn-jnl}

\usepackage{graphicx}%
\usepackage{multirow}%
\usepackage{amsmath,amssymb,amsfonts}%
\usepackage{amsthm}%
\usepackage{mathrsfs}%
\usepackage[title]{appendix}%
\usepackage{xcolor}%
\usepackage{textcomp}%
\usepackage{manyfoot}%
\usepackage{booktabs}%
\usepackage{algorithm}%
\usepackage{algorithmicx}%
\usepackage{algpseudocode}%
\usepackage{listings}%
\usepackage{natbib}
\usepackage{array}
\usepackage{algpseudocode}
\usepackage{tikz}
\providecommand{\Description}[1]{}

\theoremstyle{thmstyleone}%
\theoremstyle{thmstyletwo}%

\theoremstyle{thmstylethree}%

\begin{document}

\title[AI Washing and the Erosion of Digital Legitimacy: A Socio-Technical Perspective on Responsible Artificial Intelligence in Business]{AI Washing and the Erosion of Digital Legitimacy: A Socio-Technical Perspective on Responsible Artificial Intelligence in Business}


\author{\fnm{Nelly} \sur{Elsayed}}\email{elsayeny@ucmail.uc.edu}



\affil*[1]{\orgdiv{School of Information Technology}, \orgname{University of Cincinnati}, \orgaddress{\city{Cincinnati}, \state{Ohio}, \country{United States}}}




\abstract{The rapid evolution of artificial intelligence (AI) systems, tools, and technologies has opened up novel, unprecedented opportunities for businesses to innovate, differentiate, and compete. However, growing concerns have emerged about the use of AI in businesses, particularly AI washing, in which firms exaggerate, misrepresent, or superficially signal their AI capabilities to gain financial and reputational advantages. This paper aims to establish a conceptual foundation for understanding AI washing. In this paper, we draw on analogies from greenwashing and insights from Information Systems (IS) research on ethics, trust, signaling, and digital innovation. This paper proposes a typology of AI washing practices across four primary domains: marketing and branding, technical capability inflation, strategic signaling, and governance-based washing. In addition, we examine their organizational, industry, and societal impacts. Our investigation and analysis reveal how AI washing can lead to short-term gains; however, it also proposes severe long-term consequences, including reputational damage, erosion of trust, and misallocation of resources. Moreover, this paper examines current research directions and open questions aimed at mitigating AI washing practices and enhancing the trust and reliability of legitimate AI systems and technologies.}




\maketitle

\section{Introduction}\label{sec:Intro}

Technology advancements have created revolutionary progress enabling new forms of digital interaction and organizational transformation in all sectors, including business~\citep{ramachandran2022assessing,rymarczyk2020technologies,shugan2004impact,gorman2008evaluating}. Online shopping has increased sales and provided larger opportunities for expanding current sales~\citep{chintala2024browsing,luo2025stopping}, starting new startups~\citep{roach2024can,hsu2025remote,santisteban2021critical}, and trading between various locations and areas~\citep{eaton2002technology,baviera2012trade}. Moreover, it is expanding the services of logistics and shipping carriers~\citep{melnyk2023focused,kolasinska2022smart,lai2020impact}. Digital platforms have thus become central to business models worldwide, reshaping how firms create, deliver, and capture value~\citep{wang2025artificial,sikder2023power,zhu2010let,zhou2007online,alzoubi2022effect}.

Artificial Intelligence (AI) has been adopted in several businesses and systems due to the significant impact of AI on the digital transformation, prediction, personalization, decision making, and automation of systems, which can significantly enhance the business performance~\citep{bharadiya2023impact,callier2023ai,ransbotham2018artificial,enholm2022artificial}. On the other hand, its perceived value is vulnerable to manipulation as employing AI has not always been accompanied by transparency, accountability, or technical rigor~\citep{puchakayala2022responsible,agrawal2024accountability,tripathi2025ethical,jobin2019global,vazques2024artificial}.

There is a significant growth in the number of organizations that are engaging in AI washing, which is the practice of exaggerating, misrepresenting, or falsely claiming AI capabilities in products, services, or processes to gain reputational or financial advantage~\citep{parikh2024artificial,leffrang2023ai,al2024ai}.
AI washing phenomenon mirrors earlier corporate practices such as greenwashing, in which companies practice of making misleading or false claims about a product, service, or company’s environmental benefits to appear more eco-friendly than they are to attract environmentally conscious consumers, investors, and regulatory favor without making substantial sustainability efforts~\citep{yang2020greenwashing,miller2017greenwashing,szabo2021perceived,akturan2018does,parikh2024artificial,leffrang2023ai,al2024ai}. While such practices may offer short-term legitimacy benefits, they ultimately erode stakeholder trust, undermine ethical AI development, and distort innovation incentives~\citep{marabelli2025artificial}.

Despite its growing prevalence, AI washing remains theoretically underexplored in the field of Information Systems (IS). The current existing research primarily focuses on ethical AI use~\citep{schlagwein2023chatgpt,currie2025rethinking,eitel2021beyond,bankins2021ethical}, governance~\citep{doi:10.1177/02683962231176842,dafoe2018ai,papagiannidis2023toward}, and accountability~\citep{currie2025rethinking,vannuccini2024artificial,raja2023ai,novelli2024accountability}. However, few studies conceptualize how AI washing emerges as a socio-technical legitimacy strategy, one that fuses technological claims with symbolic signaling to manage stakeholder perceptions. Thus, addressing this gap is crucial as organizations increasingly employ AI systems and tools in their digital strategies, shaping how technological legitimacy is constructed.

This paper investigates AI washing as a distinct socio-technical phenomenon that extends beyond deceptive communication to reflect the interaction between digital innovation, legitimacy construction, and organizational strategy. Thus, this paper aims to address the following research questions:

\textit{\textbf{RQ1:} How can AI washing be conceptually differentiated from analogous forms of corporate misrepresentation such as greenwashing, and what socio-technical mechanisms underlie its emergence in organizational contexts?\\
\textbf{RQ2:} What are the implications of AI washing for organizational legitimacy, trust, and digital strategy in the context of digital transformation?} 

\noindent

This paper contributes to IS research by conceptualizing AI washing as a special form of digital misrepresentation rooted in socio-technical dynamics. The paper compares AI washing to traditional greenwashing, develops a framework for its classification and understanding, and situates it within key IS domains, including digital innovation, ethics, and organizational trust. The goal is to advance a clear theoretical understanding of how AI washing emerges, functions, and impacts organizational legitimacy in the evolving landscape of AI-driven digital transformation.

\section{Background and Related Work}\label{sec:Background}

Innovation remains a cornerstone of business competitiveness, driving efficiency, enhancing customer experiences, and maximizing profitability~\citep{mithas2016research}. Adopting cutting-edge technologies is often seen as a strategic necessity for gaining a competitive edge, and attracting both customers and investors has become the current trend~\citep{ameen2021consumer,thomas2006direct}. Digital transformation and technology adoption have accelerated this trend, with organizations embedding advanced technologies across operations, decision-making, marketing, supply chains, and customer service~\citep{abd2025advanced,omol2024organizational,westerman2014leading}.

Technology adoption has opened vast opportunities. However, it has produced multiple complex challenges~\citep{ejiaku2014technology,al2023role,faruque2024technology}. Many technological innovations exhibit a dual nature that simultaneously enables progress and poses risks to ethical, social, and governance principles~\citep{pustovit2010philosophical}. For instance, significant advancements in cloud computing and big data have led to substantial transformations across industries, prompting increased concerns about privacy, fairness, and accountability~\citep{poudel2024impact,gawankar2024anticipating,ionescu2025adopting,yang2017big}. As organizations have adopted these technologies, the tension between maximizing performance and upholding ethical integrity has become increasingly apparent~\citep{collins2009essentials,zadek1998balancing,ateeq2024impact,schweitzer2024artificial}.

Similarly, with the current evolution in AI systems, automated decision-making, insight generation, and service personalization have become very attractive to businesses and a core element of modern business strategies. Nevertheless, there have been many sparked debates on the transparency, accountability, bias, and trustworthiness of employing such systems~\citep{meduri2025accountability,ganesan2020balancing,cheong2024transparency,fu2020artificial,fu2022fair}. However, as organizations rapidly integrate AI, questions have emerged around transparency, bias, accountability, and trustworthiness~\citep{emma2024ethical,akinrinola2024navigating}. These concerns extend beyond technological functionality to include the symbolic and strategic ways in which AI is represented and promoted. How firms communicate their AI capabilities shapes stakeholder trust, investor confidence, and regulatory attention~\citep{fu2020artificial,fu2022fair}.

Adoption of AI tools and systems has been accompanied by significant promotion in both the media and corporate records~\citep{lakshika2024evolving,metha2025ai}. Businesses often highlight AI capabilities to signal innovation and technological leadership~\citep{lee2019emerging,jorzik2023artificial,mitrache2024influence}. Many organizations invest significantly in AI-driven transformation to enhance their performance. Other organizations engage in AI washing to gain a market advantage~\citep{fioravante2024beyond,al2024ai,ozturkcan2025responsible}. 

Within Information Systems (IS) research, several streams provide relevant foundations for examining this issue. There are several crucial Information Systems (IS) literature that have investigated multiple topics on digital ethics and responsible innovation, highlighting the need for transparent and accountable technology practices~\citep{mccarthy2025socio,Susarla,alsulami2025digital,cui2025exploring}.
These studies provided valuable insights on signaling theory and technology legitimacy, explaining how organizations use symbolic cues to influence perceptions of technological capability and trustworthiness~\citep{Susarla}. In addition, the literature on trust in digital technologies underscores how unverified claims or hype cycles can undermine stakeholder confidence, distort adoption decisions, and trigger regulatory backlash~\citep{mccarthy2025socio}.

Research on signaling theory and technology legitimacy explains how organizations use symbolic cues to influence perceptions of technological capability and trustworthiness~\citep{Susarla}. In addition, the literature on trust in digital technologies underscores how unverified claims or hype cycles can undermine stakeholder confidence, distort adoption decisions, and trigger regulatory backlash~\citep{mccarthy2025socio}. 

Building on this literature and studies, we conceptualize AI washing as a socio-technical legitimacy phenomenon: a strategic attempt to construct or maintain digital legitimacy through symbolic AI claims that are only weakly supported or entirely unsupported by actual technical capabilities. Understanding AI washing through this lens allows IS researchers to analyze how symbolic representations of technology shape organizational legitimacy, stakeholder trust, and the broader dynamics of digital transformation.

\subsection{Greenwashing as a Conceptual Precursor to AI Washing}
False information and misleading claims are not a novel practice~\citep{lazer2018science,petratos2023fake,guo2023debunking}. While the term "greenwashing" was coined in the 1980s, the practice of using environmental or sustainable claims to mask less desirable practices has historical precedents~\citep{de2020concepts}. According to Fisher et al.~\citep{fisher2023ancient}, Roman practices share similarities with modern greenwashing in gastronomic products and engineering processes.

Greenwashing is defined as the practice of making misleading or false claims about the environmental friendliness of an organization's products or operations to attract investors, gain regulatory favor, and enhance public perception \citep{hanny2023climate,parguel2011sustainability,pizzetti2021firms}. While such schemes may deliver short-term reputational or financial benefits, they often generate long-term risks, such as consumer backlash, regulatory penalties, loss of trust, and reputational damage, which can erode market value and stakeholder confidence \citep{wu2020bad}.

Greenwashing is defined as the practice of making misleading or false claims about the environmental friendliness of an organization's products or operations to attract investors, gain regulatory favor, and improve public perception for financial gain~\citep{hanny2023climate,parguel2011sustainability,pizzetti2021firms}. While a greenwashing scheme may deliver reputational or financial gains, it often lacks genuine environmental benefits. Such gains are short-term and can pose significant long-term consequences to businesses and stakeholders. Companies that engage in greenwashing generate long-term risks, including consumer backlash, regulatory penalties, loss of trust, and reputational damage, which can erode market value and stakeholder confidence~\citep{wu2020bad}.

Beyond reputational harm, greenwashing also distorts markets by misdirecting investments toward firms with inflated sustainability credentials, undermining the credibility of genuine sustainability efforts. These distortions have prompted regulatory responses and growing demands for transparent environmental, social, and governance (ESG) disclosures~\citep{oncioiu2020role,yu2018environmental,alsayegh2020corporate}.

The greenwashing phenomenon is analogous to providing a useful conceptual precursor to AI washing. AI washing is the most recent technological phenomenon where organizations misrepresent their use of artificial intelligence to capitalize on the hype surrounding the technology~\citep{ozturkcan2025responsible,schultz2022digital}. AI washing poses ethical and economic risks, as does greenwashing. In addition, AI washing risks compromising technological integrity, weakening stakeholder confidence, and potentially harming innovation ecosystems by diverting attention and resources away from AI-driven solutions. Drawing on the study of greenwashing helps conceptualize AI washing as a legitimacy-seeking behavior with reputational, strategic, and systemic consequences.

\subsection{Conceptual Framework: Defining and Categorizing AI Washing}

AI washing refers to the practice of exaggerating, misleading, or falsely advertising the use of AI technologies, algorithms, or systems in organizational products, services, or strategies \citep{fioravante2024beyond,ozturkcan2025responsible}. It typically involves representing a system as more autonomous, intelligent, or innovative than it truly is. The primary motivation is to gain a competitive advantage, attract investment, or build brand legitimacy by signaling alignment with the digital transformation movement.

In legitimate implementations, AI typically combines machine-learning algorithms, data pipelines, human oversight, and iterative refinement~\citep{rane2025data,harbi2023responsible,kumar2024black,loeza2025machine,high2024optimizing}. In contrast, AI washing replaces substantive innovation with symbolic adoption: the appearance of AI competence becomes a marketing or reputational device. It thus represents a form of strategic deception, where the signal of innovation is prioritized over actual technical capability.

\begin{table*}[h]
\centering
\renewcommand{\arraystretch}{1.3}
\caption{A comparison between the AI washing and greenwashing}
\label{copmare1}
\begin{tabular}{|p{2.5cm}|p{4.6cm}|p{4.6cm}|}
\hline
\textbf{Compare} &  \textbf{AI Washing} & \textbf{Greenwashing} \\
\hline
Domain &	Digital technology& Environmental sustainability\\
\hline
 Intent&	Appear innovative or AI-enabled& Appear eco-friendly\\
\hline
Negative Impacts&	Innovation failure, reputational damage
& Distrust, regulation, lawsuits \\
\hline
 Countermeasures& AI audits, explainability, IS research&Certifications, audits, transparency \\
\hline
 Primary Stakeholders&Investors, customers, partners&Investors, consumers, regulators\\
\hline
\end{tabular}
\end{table*}

AI washing is a misrepresentation phenomenon that parallels the earlier practice of greenwashing, in which companies exaggerated their environmental credentials to attract sustainability-focused consumers and investors. However, the AI washing used multiple AI tools and systems to produce deepfake content aimed at the same target. Table~\ref{copmare1} shows a comparison between the AI washing and greenwashing at organizations. Both AI washing and greenwashing exploit an information asymmetry between companies and their stakeholders, where technical claims are challenging to verify externally.

Both greenwashing and AI washing phenomena exploit information asymmetries between organizations and their stakeholders, where technical or environmental claims are challenging to verify externally. To further analyze this analogy, we propose a typology of AI-washing practices inspired by the greenwashing literature \citep{burbano2016social,delmas2011drivers,floridi2022unified,mittelstadt2016ethics,ransbotham2017reshaping,lyon2015means,seele2017greenwashing,zerilli2019transparency,connelly37signaling,von2018artificial,vial2021understanding,gal2022research,COUSSEMENT2024114276}.

The topology of AI washing practices and their corresponding greenwashing analogs is presented in Table~\ref{typology}, highlighting the similarities between greenwashing and AI washing along both intentional deception and misleading communication axes.

\begin{sidewaystable}
\centering
\caption{Topology of AI Washing with Greenwashing Analogs}
\label{typology}
\begin{tabular*}{\textheight}{@{\extracolsep\fill}p{2.5cm}p{4.2cm}p{5cm}p{5cm}}
\toprule
\textbf{Type} & \textbf{Description} & \textbf{AI Washing Example} & \textbf{Analog Greenwashing Example} \\
\midrule
Fictitious AI Claims & 
Claiming AI is used when it is not used at all & 
Re-branding a rules-based workflow as ``AI-powered'' & 
Marketing a product as ``100\% eco-friendly'' with no certification. \\

Overstated Autonomy/Capability & 
Exaggerating intelligence, autonomy, or accuracy to signal innovation & 
Human-in-the-loop service advertised as fully autonomous with ``99\% accuracy'' & 
Calling a hybrid car ``zero-emissions''. \\

Vague or Misleading Terminology & 
Using popular AI terminology without specifying models, data, or scope & 
Using ``intelligent,'' ``smart,'' ``machine learning,'' or ``deep learning'' with no technical detail & 
Using ``green'' or ``natural'' with no criteria. \\

Symbolic Implementation (Tokenism) & 
Adding superficial features to justify AI branding without real value-add & 
Attaching a basic pre-trained model that does not change outcomes & 
Adding a small ``recycled'' label while the core process remains polluting. \\

Omitted Limitations & 
Concealing risks, human oversight, or known failures & 
Not disclosing manual review of outputs or bias from training data & 
Not disclosing upstream emissions or energy intensity. \\

Ethics/Trustwashing & 
Promoting Responsible AI without governance or controls & 
Publishing an AI ethics pledge with no policy, board, or audits & 
Issuing sustainability pledges while practices remain unchanged. \\
\botrule
\end{tabular*}
\end{sidewaystable}

\begin{figure}[t]
    \centering
    \includegraphics[width=8cm, height= 8.5 cm]{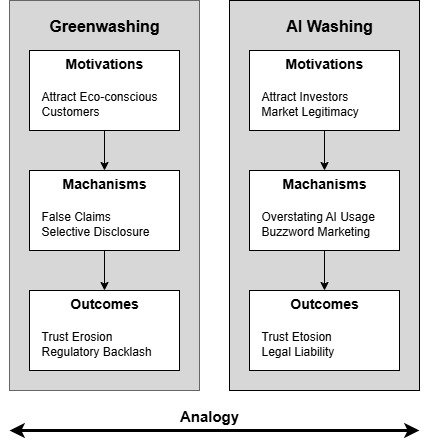}
    \caption{The AI washing within the broader context of deceptive corporate practices, using greenwashing as the historical reference point.}
    \label{concept}
\end{figure}

The comparison between greenwashing and AI washing highlights how organizations may exploit emergent technologies and social concerns for reputational advantage. Both practices involve signaling strategies that may mislead stakeholders and undermine long-term trust if discovered. To further clarify this analogy, we present a conceptual framework (Figure~\ref{concept}) that situates AI washing within the broader context of deceptive corporate practices, using greenwashing as a historical reference point.

The comparisons in Table~\ref{copmare1} and Table~\ref{typology} show how both practices use symbolic signaling to exploit emerging technological or social narratives for reputational gain. While greenwashing appeals to environmental virtue, AI washing leverages the cultural capital of artificial intelligence to signal innovation. Both, however, carry serious long-term consequences: when uncovered, they erode trust, distort investment, and compromise institutional credibility.

Greenwashing and AI washing practices demonstrate how emerging technologies and social concerns can be leveraged by various organizations for reputation benefits and financial gain. Greenwashing relied on overstated environmental claims. AI washing leverages the current hype surrounding artificial intelligence technologies and systems. 
Both greenwashing and AI washing practices involve signaling strategies that may mislead stakeholders. In addition, both practices lead to severe long-term consequences, including trust issues if discovered. We present a conceptual framework that clarifies this analogy by situating AI washing within the broader context of deceptive corporate practices, using greenwashing as a historical reference point.

Figure~\ref{concept} illustrates the conceptual positioning of AI washing within the broader landscape of deceptive corporate practices. Drawing on greenwashing as a historical reference point, the figure situates AI washing as a digitally enabled form of legitimacy manipulation that operates across socio-technical dimensions, technological, organizational, and communicative. By highlighting this relationship, the framework underscores how symbolic technological claims can serve as instruments of reputation management and legitimacy construction, even in the absence of substantive innovation. This conceptual foundation serves as the basis for the next section, which distinguishes AI washing from greenwashing and develops a socio-technical understanding of digital legitimacy.

\section{Conceptual Differentiation Between AI Washing and Greenwashing: Toward a Socio-Technical Understanding of Digital Legitimacy}

While AI washing and greenwashing share a foundation in symbolic misrepresentation, they diverge markedly in their underlying logics, socio-technical structures, and implications for organizational legitimacy~\citep{leffrang2023ai}. Greenwashing originated within environmental and sustainability discourses that sought to align corporate behavior with ecological responsibility~\citep{marciniak2009greenwashing}. In contrast, AI washing emerges from digital innovations and transformations in which artificial intelligence functions as both a technical infrastructure and a symbol of modernity, serving as a strategic narrative of innovation~\citep{duer2013applications,van2024big,lavrentyeva2019artificial,malik2022digital,kitsios2021artificial,oyekunle2024digital,ayoko2021digital,fioravante2024beyond}.

Understanding these distinctions requires moving beyond analogy to examine how digital technologies reshape the mechanisms of signaling, verification, and trust. Building on the conceptual groundwork in Section~\ref{sec:Background}, this section develops a socio-technical framework for differentiating AI washing from its greenwashing predecessor, focusing on how organizations construct and negotiate digital legitimacy in technologically mediated environments.

\subsection{From Environmental to Digital Legitimacy}
In greenwashing, legitimacy is primarily moral and environmental—organizations seek societal approval by signaling environmental responsibility or sustainability commitment~\citep{de2020concepts}. Legitimacy derives from the alignment between a firm's environmental claims and collective expectations concerning ethical resource use, ecological preservation, and global sustainability goals.

In contrast, AI washing centers on technical and symbolic legitimacy. Firms aim to secure digital legitimacy—the perception that they possess advanced technological capability, data competence, and innovation leadership. In this context, legitimacy is co-constructed not only through corporate communication but also through algorithmic systems, data infrastructures, and human–machine interactions that perform and reinforce a firm's technological credibility.

Environmental legitimacy can be externally verified through audits or certifications. However, digital legitimacy is epistemically unclear. The technical complexity of AI systems, algorithms, and restricted data processes creates information asymmetries between organizational claims and stakeholder observation~\citep{lu2020algorithmic,rezaei2024ai,matthews2025review,han2022impact}. This uncertainty heightens the potential for AI washing, as verification requires both technical expertise and access to internal systems, which are rarely available information to consumers, regulators, or investors.

\subsection{Mechanisms of Misrepresentation: Symbolic, Technical, and Organizational Dimensions}

The mechanisms of AI washing differ from those of greenwashing in both form and complexity. Greenwashing typically operates through symbolic disclosure, such as eco-labels, sustainability reports, or marketing narratives that exaggerate environmental practices. In contrast, AI washing involves technical and organizational misrepresentation, blending narrative, design, and data manipulation.

These intertwined technical and organizational mechanisms create diverse expressions of AI washing within firms. Depending on how narrative, design, and governance elements interact, AI washing may appear in varying degrees of sophistication and intent. To better understand these variations, three primary dimensions can be distinguished:

\begin{itemize}
    \item \textbf{\textit{Symbolic AI washing}} occurs when firms use AI terminology without an actual implementation. 
    \item \textit{\textbf{Technical AI washing}} emerges when companies deploy limited or basic algorithms while present them as advanced AI systems.
    \item \textit{\textbf{Organizational AI washing}} occurs when companies overstate their internal AI capacity such as claiming to have "AI teams" or "AI research department" that serve marketing rather than technical functions.
\end{itemize}

These tripartite dimensions demonstrate that AI washing operates across the organizational stack, shaping not only external communication but also internal decision-making and investment priorities. It highlights how misrepresentation becomes embedded within the socio-technical configuration of organizations rather than confined to surface-level rhetoric.

\subsection{Mechanisms of Legitimacy Construction}
AI washing and greenwashing also differ in how legitimacy is constructed and maintained.
Greenwashing primarily relies on symbolic compliance mechanisms, such as sustainability reporting, eco-labels, and corporate social responsibility disclosures. These mechanisms operate within institutionalized frameworks, where environmental indicators and third-party audits provide some accountability and verification. In contrast, AI washing depends on technical opacity and innovation rhetoric. Legitimacy is achieved through signaling practices that emphasize complexity, futurism, and technological inevitability~\citep{bjornali2017reveal,vanderborght2025visions}. Terms such as intelligent, smart, autonomous, or AI-enabled prompt progress while obscuring substance~\citep{verma2024relevance}. Organizations often rely on performative artifacts~\citep{glaser2017design}, such as prototypes, dashboards, or conversational agents, that visually or interactively demonstrate "intelligence," even when underlying systems remain algorithmic-based or manually operated.

Consequently, legitimacy in AI washing becomes a technologically mediated construct, shaped by a combined narrative framing and system design. This process aligns with the concept of digital performativity, where the representation of capability itself generates trust, investment, and adoption, independent of technical authenticity.

\subsection{A Socio-Technical Framework of Digital Legitimacy}
AI washing demonstrates how legitimacy in digital contexts is co-produced through interactions between technological infrastructures, organizational strategies, and social interpretations. Unlike earlier forms of symbolic misrepresentation, AI washing operates within systems characterized by algorithmic opacity, rapid innovation cycles, and high stakeholder uncertainty.

\begin{figure}[t]
    \centering
    \includegraphics[width=6.5cm, height= 6 cm]{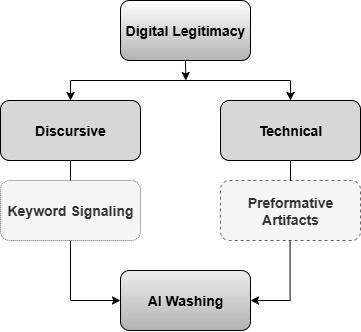}
    \caption{A socio-technical framework of digital legitimacy and AI washing.}
    \Description{A conceptual diagram showing how AI washing operates across symbolic, technical, and organizational dimensions. Arrows illustrate relationships between legitimacy mechanisms, firm actions, and stakeholder perception.}
    \label{framework}
\end{figure}

This subsection introduces a socio-technical framework of digital legitimacy that situates AI washing as a distinct manifestation of legitimacy-seeking behavior in the age of intelligent technologies (Figure~\ref{framework}). The framework conceptualizes AI washing as the outcome of three intersecting dynamics:

\begin{itemize}
    \item \textit{\textbf{Technological opacity:}} Limited stakeholder understanding of AI processes creates space for unverifiable claims.
\item \textit{\textbf{Rhetorical innovation:}} Organizations construct narratives of intelligence and autonomy to align with the discourse of digital transformation.

\item \textit{\textbf{Institutional legitimacy pressures:}} Firms face growing expectations from investors, customers, and policymakers to demonstrate AI capability, regardless of actual deployment depth.
\end{itemize}

Together, these dynamics generate a feedback loop in which technological hype becomes self-reinforcing: performative claims attract attention and resources, which in turn sustain the legitimacy of the AI narrative.

Building upon these insights, we propose a socio-technical model that conceptualizes how digital legitimacy is constructed through both discursive and technical mechanisms. This model highlights how organizational narratives (keyword signaling) and technological displays (performative artifacts) jointly reinforce the perception of AI capability, thereby enabling AI washing. As shown in Figure~\ref{framework}, digital legitimacy functions as the central mediator between organizational signaling practices and stakeholder perception, illustrating how AI washing emerges from the interplay between technological opacity and discursive framing.

\subsection{The Socio-Technical Nature of AI Washing}
AI washing highlights the socio-technical negotiation between technological claims and social validation~\citep{holton2021people,lu2009socio,flichy2007understanding}. AI washing manipulates epistemic authority where the perceived expertise and credibility of organizations in the digital domain. Stakeholders, investors, and consumers cannot frequently assess algorithmic authenticity, allowing firms to construct legitimacy through opacity.

This dynamic creates a digital legitimacy paradox: the more technically complex AI systems become, the harder they are to verify, thereby increasing opportunities for symbolic signaling~\citep{dobbe2021hard,von2021transparency,brundage2020toward}. Thus, AI washing exposes how digital transformation amplifies asymmetries of knowledge and power between firms and their stakeholders.

In this context, legitimacy is not merely reputational but performative~\citep{martin2023algorithmic,krahmann2017legitimizing}. Organizations build "AI capability" to secure resources, build trust, and differentiate in digital markets. This aligns with theories of technological performativity, which hold that representations of innovation shape economic and institutional realities.


\section{AI Adoption in Business Contexts}

AI technologies and AI-based systems have been widely adopted across diverse business sectors~\citep{sandeep2022understand,george2023review,bharadiya2023rise}, including finance~\citep{cao2022ai,jafar2023ai,aleksandrova2023survey}, marketing~\citep{mariani2022ai,chintalapati2022artificial,kumar2024ethical}, healthcare~\citep{chinta2025ai,sullivan2025can,tekkecsin2019artificial,bilal2022role,rong2020artificial,muyskens2024can}, retail~\citep{oosthuizen2021artificial,heins2023artificial,cao2021artificial}, and logistics~\citep{chen2024artificial,richey2023artificial,sohrabi2023artificial,pandian2019artificial}. This broad adoption stems from the significant strategic advantages that AI offers to organizations, such as predictive analytics, customer personalization, decision support, and process automation. The accessibility of AI has further expanded through open-source libraries and a growing ecosystem of cloud-based services, making it easier than ever to integrate AI capabilities into business processes and enterprise applications~\citep{langenkamp2022open,naveen2024overview,lee2018ai,okwu2024future,bagchi2020conceptualising}.

From the information systems (IS) perspective, AI represents a transformative capability that reshapes the structure and dynamics of organizations~\citep{dennehy2023artificial,dwivedi2021artificial,liu2022trustworthy}. AI can influence how decisions are made, how customers interact with firms, and how businesses perceive innovation itself. However, a new form of institutional pressure has emerged, including investors, partners, and stakeholders who increasingly expect organizations to demonstrate AI adoption from this perspective to remain competitive and relevant to the current technology trends. As a result, many firms aim to signal technological sophistication by claiming AI capabilities regardless of whether those capabilities are substantive or implemented in practice.

There are clear advantages to AI adoption in business. These include competitive differentiation, operational efficiency gains, fostering data-driven innovation, accelerating processes, and improving overall performance~\citep{maiti2025study,chan2025application,akbarighatar2025enacting}. However, these advantages exist alongside risks exacerbated by the current AI hype cycle. Over-investment in underperforming systems, strategic misalignment between AI capabilities and organizational goals, and growing ethical and governance challenges are increasingly observed across industries~\citep{jobin2019global,boncella2024ai,rana2022understanding}.

In addition, limited explainability, embedded bias, and consumer confusion when AI claims fail to meet expectations create reputational and regulatory vulnerabilities~\citep{sicular2020hype,oosterhoff2020artificial,vashisth2019hype,zimmerli2025artificial,jazairy2025impact}.

This duality between opportunity and exaggeration underscores the need for both researchers and practitioners to evaluate how AI is integrated, communicated, and legitimized. The institutional drive to appear "AI-enabled" often leads to symbolic representations of intelligence rather than genuine innovation, paving the way for AI washing practices within the business ecosystem.

\section{Types of AI Washing in Business}

AI washing exemplifies several distinct forms, reflecting the different ways firms exaggerate, misrepresent, or strategically frame their use of AI technologies. These forms vary in scope, intent, and organizational depth from superficial marketing language to systemic misrepresentation of governance practices. The four general categories can be identified as:

\begin{itemize}

    \item \textit{Marketing and Branding AI Washing:} In this form of AI washing, firms use promotional campaigns, labels, or advertising to present their offerings as "AI-powered," regardless of whether AI technologies are actually embedded or not~\citep{haridasan2025ai}. The primary objective of this type of AI washing is to capitalize on AI's positive connotations to capture consumer attention and achieve competitive differentiation.

    \item \textit{Technical Capability Inflation:} In this type of AI washing, organizations overstate the usage and embedding of intelligent solutions in their AI systems. This form of AI washing relies on partial truth~\citep{anand2025ignoble}s. For example, the system may incorporate some machine learning (ML) elements but is presented as far more capable than it actually is. 
    
    \item \textit{Strategic AI Signaling:} This type of AI washing occurs when businesses announce AI initiatives, partnerships, or research and development (R\&D) projects more for their symbolic value rather than for substantive development~\citep{fioravante2024beyond}. Similar practices occur in greenwashing, where firms join sustainability pledges or publicize environmental memberships without taking meaningful action.

    \item \textit{Governance and Ethical AI Washing:} This type of AI washing overstates commitments to the responsible use of AI, where responsible AI is based on building and deploying artificial intelligence systems ethically, with human well-being at the forefront~\citep{bietti2021ethics}. This prioritizes fairness, transparency, accountability, privacy, safety, and human oversight to prevent harm, bias, and misuse. Companies may publish AI ethics principles or declare their implementation of fairness and transparency frameworks without actually implementing governance processes, measurable practices, or mechanisms to ensure compliance. This form is considered the most harmful, as it exploits ethical concerns for legitimacy while doing little to mitigate actual risks. 
\end{itemize}

\begin{sidewaystable}
\centering
\caption{Types of AI Washing in Business Contexts}
\label{AiWashing_types}
\begin{tabular*}{\textheight}{@{\extracolsep\fill}p{3cm}p{5cm}p{5cm}p{5cm}}
\toprule
\textbf{Type of AI Washing} & \textbf{Description} & \textbf{Primary Goal} & \textbf{Potential Consequences} \\
\midrule

\textit{Marketing and Branding AI Washing} & 
Firms promote products or services as ``AI-powered'' in advertising or product labels, even when little to no AI technology is used. & 
To attract consumers, investors, and partners by leveraging the hype around AI and enhancing perceived innovation. & 
Erosion of consumer trust, market misinformation, and misallocation of resources toward non-AI solutions. \\

\midrule

\textit{Technical Capability Inflation} & 
Organizations exaggerate or selectively present the technical sophistication of AI systems. Basic automation or rule-based tools may be described as advanced AI. & 
To signal technical advancement, justify higher pricing, or appeal to technologically driven investors. & 
Reduced transparency, internal misalignment, and eventual reputational damage when capabilities fail to meet expectations. \\

\midrule

\textit{Strategic AI Signaling} & 
Firms publicize AI initiatives, partnerships, or R\&D projects mainly for symbolic value or legitimacy rather than substantive development. & 
To appear technologically progressive and sustain investor confidence without major financial or technical investment. & 
Misleading investors and stakeholders, creating inflated market valuations and strategic misdirection. \\

\midrule

\textit{Governance and Ethical AI Washing} & 
Companies publicly declare AI ethics principles or responsible AI practices without implementing enforceable measures or transparent accountability mechanisms. & 
To gain legitimacy and public approval by appearing socially responsible and aligned with ethical AI norms. & 
Loss of stakeholder trust, regulatory scrutiny, and harm to users if ethical risks are not mitigated in practice. \\
\botrule
\end{tabular*}
\end{sidewaystable}

Table~\ref{AiWashing_types} presents a typology of AI washing practices that occur across different layers of organizational activity. Each category reflects a distinct mechanism through which firms strategically construct or exaggerate AI-related legitimacy. While marketing and branding AI washing focuses on shaping consumer perception through symbolic association with innovation, technical capability inflation, and strategic AI signaling operate at deeper organizational and infrastructural levels, influencing investment, development priorities, and stakeholder relations. Governance and ethical AI washing, the most insidious form, manipulates the language of responsible AI to secure trust and compliance legitimacy without embedding substantive accountability mechanisms. Together, these categories illustrate how AI washing extends beyond mere promotional exaggeration to encompass systemic practices that shape technological narratives, institutional credibility, and digital ethics in contemporary business contexts.

Together, these four categories reveal that AI washing is not a single deceptive act but a multi-layered institutional phenomenon that operates across marketing, technology, strategy, and governance.  

\section{Insights from IS Research on Ethics, Trust, and Innovation}

The parallels between greenwashing and AI washing highlight the importance of examining how organizations present and communicate their technological practices~\citep{lumineau2025roadmap}. There are several Information Systems (IS) field research investigations that provide valuable theoretical lenses for understanding greenwashing and AI washing from the perspectives of digital ethics, digital system research, and signaling theory.~\citet{markus2017datification} studied how the adoption of emerging technologies outpaces ethical governance. In the trust in digital systems research, Gefen et al.~\citep{gefen2008research} investigated how misrepresentation can erode both customer and stakeholder confidence and the resulting consequences for organizational legitimacy.
~\citet{connelly37signaling} investigated how organizations strategically communicate innovation to differentiate themselves in competitive markets.~\citet{lyytinen2016digital} study showed that organizations often adopt emerging technologies' symbolic value in addition to the efficiency gains.

By bringing these streams of IS literature together, we can better situate AI washing as both a continuation of earlier challenges in technological adoption and as a novel form of misrepresentation with unique implications for the AI-driven economy. In this respect, the challenges of AI washing resonate strongly with three established streams of IS research: ethics in technology use, trust in information systems and organizations, and digital innovation signaling.

\begin{itemize}
\item \textit{\textbf{Ethics and Responsible Innovation in IS:}} IS research has consistently determined that the adoption of new technologies is not value-neutral~\citep{orlikowski2016digital}. Several studies have examined digital ethics, showing that new technologies pose dangers, including over-promising technological capabilities, uncertainty about how algorithms make decisions, and exaggerated corporate claims about digital transformation initiatives~\citep{martin2019ethical,culnan2009ethics,charton2025operationalising}. AI washing reflects a broader problem of how firms strategically use narratives to gain legitimacy while sidelining ethical responsibilities. While AI washing undermines by prioritizing market gains over stakeholder trust, responsible innovation frameworks in IS emphasize the urgent need for accountability, transparency, and alignment with societal values and principles that AI washing undermines. Thus, addressing AI washing is not just about ethics, but a critical need for the future of responsible digital innovation.

\item \textit{\textbf{Trust as a Central Mechanism:}} Trust is one of the primary concepts in Information Systems (IS) research, serving as an enabler and an outcome of digital adoption~\citep{Gefen2003,Pavlou2003,muller2024adopting,steininger2019linking}. User trust has been identified as the foundation for sustained engagement in e-commerce, online platforms, or enterprise systems~\citep{Bhattacherjee2001}. Engaging in AI washing within any organization leads to erosion of customer and investor trust when the customer and/or investor recognizes that the AI capabilities are overstated or that the organization's claims are inflated. The erosion of trust can have consequences, including reputational and financial damage, for the organization, which can sometimes be irreversible. Thus, according to the IS literature, AI washing is not just an ethical issue but also a strategic liability for organizations' digital ecosystems.

\item \textit{\textbf{Digital Innovation, Hype, and Signaling}}   
Broadly, IS research focuses on how businesses use innovation signals to shape perceptions of technological leadership, attract resources, and establish the legitimacy of their systems~\citep{Fichman2014,Gregory2021}. 
According to~\citep{gregory2015paradoxes}, such signaling is not inherently deceptive, as it can be a rational strategy for communicating innovation readiness and competitive positioning. However, a risk of creating a legitimacy gap between external expectations and internal reality arises when signals overstate capabilities or exaggerate business progress. The phenomenon of hype cycles highlights the systemic risks associated with inflated signaling, in which new technologies are rapidly promoted, overpromised, and subsequently questioned~\citep{peltier2024artificial,urbinati2020role}. According to that, AI washing can be conceptualized as a specific form of strategic missignaling. When organizations overstating their AI capabilities, businesses attempt to ride the wave of AI enthusiasm and tap into the legitimacy associated with being "AI-driven."  While AI washing may lead to short-term benefits, such as market attention or media coverage, several IS research alerts indicate that prolonged divergence and its consequences can cause disillusionment, loss of trust, and damaged reputations. In addition, inflated AI signaling can erode institutional trust in AI as a whole, leading to field-level consequences in which stakeholders become skeptical of genuine AI innovations~\citep{meyer2023international}. Thus, AI washing extends beyond the business-level ethical challenge to pose a systemic risk to the innovation ecosystem. AI washing risks triggering a broader erosion of credibility around AI adoption, making it harder for authentic innovators to differentiate themselves. 
\end{itemize}

Based on that, AI washing positions exceed the ethical concerns. It represents a structural challenge for businesses, organizations, and information systems. The following section explores these challenges in detail, examining the direct and indirect impacts of AI washing at the organizational, industry, and system levels.

 \begin{figure}[t]
    \centering
    \includegraphics[width=12cm, height= 12 cm]{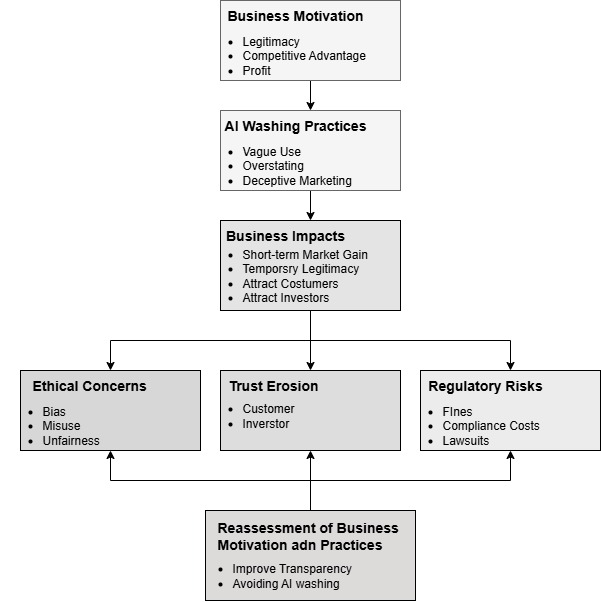}
    \caption{A dynamic framework for AI washing in business contexts.}
    \Description{The diagram shows the AI washing in business content and the interaction between these components.}
    \label{impact}
\end{figure}

\section{Challenges and Impacts of AI Washing on Businesses and Information Systems}

AI washing poses significant challenges for organizations, industries, and socio-technical systems. While overstating or misrepresenting AI capabilities can produce short-term gains—such as investor attention or enhanced market visibility—these benefits come at the expense of long-term trust, legitimacy, and innovation capacity. Figure~\ref{impact} illustrates how business motivations drive AI washing practices, generating immediate advantages but also leading to severe long-term negative consequences. Feedback loops indicate that ethical concerns, trust erosion, and regulatory risks often compel firms to reevaluate their strategies, potentially prompting either greater transparency or deeper reliance on deceptive AI narratives.

\begin{figure}[t]
    \centering
    \includegraphics[width=6cm, height=4cm]{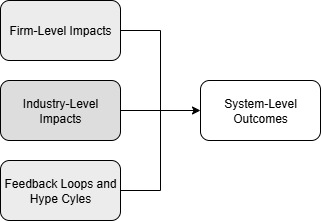}
    \caption{Multi-level impacts of AI washing on firms, industries, and socio-technical systems.}
    \Description{The diagram shows the impact of AI washing on different levels including firm, industry, and feedback loops.}
    \label{levels_impact}
\end{figure}

Although such practices may initially offer reputational or financial benefits, they also produce cascading negative consequences across multiple levels. At the firm level, organizations face reputational and compliance risks; at the industry level, distorted competition and reduced innovation; and at the system level, a broader erosion of institutional legitimacy and stakeholder trust~\citep{liang2025greenwashing}. 
These three levels—firm, industry, and system—are tightly interconnected~\citep{cheng2017measuring}. 
Figure~\ref{levels_impact} visualizes these cascading relationships, illustrating how feedback loops and hype cycles amplify the adverse impacts of AI washing. Firm-level misrepresentations aggregate into industry-wide distortions, which in turn lead to systemic consequences such as regulatory backlash and diminished public confidence.

\subsection{Firm-Level Impacts}

At the organizational level, AI washing directly undermines integrity and credibility. Companies that exaggerate AI adoption or inflate system capabilities may experience temporary boosts in valuation or customer engagement. However, once such claims are exposed, they often face severe reputational damage~\citep{mikalef2022thinking}. 
Trust erosion affects both internal and external stakeholders: customers lose confidence in digital offerings, and employees encounter ethical dilemmas that reduce morale and organizational identification. Moreover, legal and compliance risks escalate as regulators increase scrutiny over AI-related misrepresentation in marketing, product design, and governance disclosures.

\subsection{Industry-Level Impacts}

At the industry level, AI washing distorts competitive dynamics. Firms engaging in deceptive signaling gain a short-term advantage over competitors who genuinely invest in costly R\&D~\citep{krakowski2023artificial,abrardi2022artificial}. 
This creates a disincentive for authentic innovators, discouraging further technological investment and undermining industry-wide innovation trajectories. 
High-profile incidents of AI washing further contribute to reputational contagion, in which skepticism about one firm's AI claims spreads across the industry, slowing the diffusion of responsible AI practices and diminishing the credibility of the field as a whole.

\subsection{System-Level Impacts}

From a system-level perspective, AI washing contributes to cyclical patterns of technological hype and disillusionment that destabilize socio-technical ecosystems~\citep{dedehayir2016hype,linden2003understanding}. 
Inflated expectations accelerate adoption and investment during the hype phase, but when these promises fail to materialize, the resulting skepticism erodes institutional legitimacy. 
This can lead to reactive regulatory interventions and policy backlash, introducing stricter compliance frameworks that may inadvertently stifle genuine innovation. 
Over time, these cycles misallocate resources, diverting them from socially beneficial AI applications and thereby reducing the overall societal impact of technological progress.

Table~\ref{tab:Impacts} summarizes the main challenges and corresponding impacts of AI washing at the firm, industry, and system levels. 
It highlights how the effects compound across levels, transforming AI washing from an isolated organizational issue into a structural threat to digital trust and innovation ecosystems.

\begin{sidewaystable}
\centering
\caption{Challenges and Impacts of AI Washing Across Multiple Levels}
\label{tab:Impacts}
\begin{tabular*}{\textheight}{@{\extracolsep\fill}p{1cm}p{7.5cm}p{7.5cm}}
\toprule
\textbf{Level} & \textbf{Major Challenges} & \textbf{Impacts} \\
\midrule

\textbf{Firm} & 
Erosion of customer and investor trust \newline
Legal and regulatory exposure \newline
Operational inefficiencies and internal misalignment \newline
Ethical dilemmas and employee disengagement &
Loss of reputation and customer loyalty \newline
Penalties, lawsuits, and compliance risks \newline
Reduced innovation capacity \newline
High turnover and reduced morale \\

\midrule

\textbf{Industry} & 
Market distortions driven by inflated AI claims \newline
Competitive pressure to exaggerate \newline
Dilution of technical and ethical standards &
Innovation credibility gap \newline
Decline in investor confidence and stakeholder skepticism \newline
Confusion over responsible AI regulation and practices \\

\midrule

\textbf{System} & 
Erosion of institutional legitimacy \newline
Unrealistic expectations fueling hype cycles \newline
Regulatory backlash and overregulation \newline
Resource misallocation &
Public distrust in AI and related technologies \newline
Slowdown of genuine AI adoption and investment \newline
Innovation constraints and compliance burdens \newline
Reduced societal benefit of AI innovation \\
\botrule
\end{tabular*}
\end{sidewaystable}

\section{Future Directions and Research Open Questions}

\begin{figure}
    \centering
    \includegraphics[height= 4 cm, width=6cm]{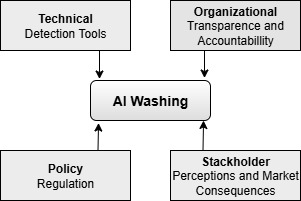}
    \caption{Conceptual overview of future research directions and open questions on AI washing.}
    \Description{A conceptual diagram showing the future research directions and open questions on AI washing.}
    \label{future_research}
\end{figure}

With the current rapid evolution of AI systems and technologies, AI washing remains an emerging phenomenon. There is significant uncertainty surrounding the use of AI systems and technologies, as well as their implications for businesses, markets, stakeholders, and the overall performance of these systems. While research on greenwashing provides a strong foundation, AI washing presents novel challenges and creates new barriers that require deeper investigations.
Figure~\ref{future_research} provides a conceptual overview of future research directions and open questions surrounding AI washing, highlighting the interplay among conceptual, regulatory, technical, organizational, and ethical dimensions that warrant further exploration.

Thus, in this paper, we investigated current directions in information systems and business that can move beyond conceptual discussions toward the development of systematic methods for detecting, preventing, and governing AI washing. Based on this analysis, the key research directions and open questions for advancing knowledge and practice in this area are as follows:

\begin{itemize}
\item \textbf{Developing Conceptual and Measurement Frameworks:}  
A significant barrier to studying AI washing is the absence of standardized definitions and measurement frameworks. Currently, there are no consistent criteria for identifying the existence or degree of AI washing within firms. This raises a fundamental research question: \textit{How can researchers objectively measure the gap between AI-related claims and the actual technological deployment within organizations?}  
Future research should prioritize developing standardized taxonomies and operational measures, such as benchmarks for claimed versus actual AI capabilities, analytical frameworks for assessing marketing content, and rubrics to evaluate responsible AI commitments.

\item \textbf{Regulation and Policy:}  
Regulations governing the responsible use of AI in business, education, and healthcare remain underdeveloped, particularly as both traditional AI and generative AI continue to evolve. Regulatory bodies must address misleading AI claims, misuse, and the need for transparency and accountability. Thus, a critical question emerges: \textit{What policy interventions can effectively reduce AI washing without hindering legitimate AI adoption across industries?}  
Answering this question requires collaboration between policymakers, researchers, and industry leaders to balance innovation with ethical governance.

\item \textbf{Technical Solutions for Detecting AI Washing:}  
Technological solutions, particularly AI-driven tools, offer promising approaches to identify AI washing at scale. This leads to the question: \textit{Can AI-based systems provide scalable and transparent mechanisms for detecting AI washing in corporate disclosures and product documentation?}  
Advances in natural language processing, auditing algorithms, and explainable AI (XAI) can be leveraged to analyze large-scale organizational data—including marketing material and system outputs—to detect inconsistencies between claims and actual system performance.

\item \textbf{Organizational Transparency and Accountability Mechanisms:}  
Organizations play a critical role in mitigating AI washing. Therefore, a key research question is: \textit{What organizational capabilities and governance mechanisms are most effective in reducing incentives for AI washing?}  
Research should explore how internal structures—such as governance frameworks, third-party audits, and ethical reporting standards—can promote transparency, accountability, and integrity in AI-related communications.

\item \textbf{Stakeholder Perceptions and Market Consequences:}  
Stakeholders’ reactions to AI washing remain underexplored. Drawing from the greenwashing literature, inflated claims may yield temporary legitimacy but erode trust when exposed. Thus, researchers should examine: \textit{How does AI washing affect consumer trust, investor decision-making, and employee morale in the long run?}  
Understanding these perceptions can motivate organizations to adopt truthful and ethical AI communication practices.

\item \textbf{Cross-Cultural and Cross-Industry Comparisons:}  
The prevalence and manifestation of AI washing may vary significantly across industries and cultural contexts. This raises an important question: \textit{How do industry characteristics and cultural norms shape the occurrence and detection of AI washing?}  
Comparative studies can reveal how different institutional settings influence firms’ incentives, regulatory compliance, and transparency efforts.

\item \textbf{Leveraging Explainability and Responsible AI:}  
Explainable AI (XAI) and responsible AI frameworks can enhance transparency and verifiability. This opens the question: \textit{How can businesses use explainable AI (XAI) to produce credible, verifiable reports and prevent exaggerated AI claims?}  
Embedding explainability into AI reporting practices can strengthen trust and reduce the occurrence of deceptive claims.

\item \textbf{Building Trustworthy AI Ecosystems:}  
Since AI washing affects multiple layers of the socio-technical landscape, multi-level strategies are necessary. A central research question here is: \textit{How can business ecosystems be designed to discourage AI washing while promoting legitimate AI innovation?}  
Addressing this question requires collaborative engagement among firms, developers, civil society, and regulators to build trustworthy, transparent, and innovation-friendly AI ecosystems.

\end{itemize}

\section{Conclusion}
This paper investigated the emerging phenomenon of AI washing as a critical challenge for businesses and information systems. Drawing on concepts and analogies from greenwashing, we argued that AI washing represents a spectrum of deceptive or overstated practices—ranging from exaggerated marketing claims to inflated technical capabilities and superficial ethical commitments. The primary goal behind these practices is often to attract investment and consumer attention, producing short-term gains at the cost of long-term trust, legitimacy, and innovation capacity. Such outcomes can have cascading consequences at the firm, industry, and societal levels, undermining both technological credibility and responsible innovation.

Our work provides a conceptual foundation for understanding AI washing by drawing on insights from IS research on ethics, trust, signaling, and digital innovation. We integrated perspectives from business practice and legitimacy theory to frame AI washing as a socio-technical legitimacy challenge in digital transformation. Furthermore, we proposed a typology of AI washing—encompassing marketing and branding, technical capability inflation, strategic signaling, and governance-based practices—which offers a systematic framework for academic inquiry and managerial assessment.

The paper contributes to Information Systems theory, business practice, and future research. For IS theory, it extends work on signaling, ethics, and digital innovation into the domain of AI washing, revealing how technological misrepresentation disrupts trust and legitimacy in digital ecosystems. For business practice, it provides a structured typology that helps managers and policymakers recognize and mitigate misleading AI-related practices. For research, it outlines key directions and open questions that can guide the development of measurable frameworks, regulatory policies, and accountability mechanisms to counter AI washing while promoting responsible AI adoption.
Collectively, these contributions establish AI washing as a distinct, consequential, and urgent topic for IS and business scholarship—one that demands rigorous conceptualization, empirical investigation, and cross-sector collaboration to safeguard the integrity of AI-driven transformation.









\section*{Declarations}

\begin{itemize}
\item \textbf{Funding:} No funding was received to assist with the preparation of this manuscript.
\item Conflict of interest/Competing interests (check journal-specific guidelines for which heading to use)
\item \textbf{Ethics approval and consent to participate: }Ethics approval and consent to participate are not applicable to this article, as no studies involving human participants or animals were performed by the authors.
\item \textbf{Consent for publication:} Not applicable. The manuscript does not contain any individual person's data in any form.
\item \textbf{Data availability:} Data sharing is not applicable to this article as no new datasets were generated or analyzed during the current study. 
\item \textbf{Materials availability:} Not applicable. No materials were used or generated in this purely theoretical study.
\item \textbf{Code availability:} Not applicable. No custom code or software was generated or used during this study.
\item \textbf{Author contribution:} The author confirms sole responsibility for study conception and design, interpretation, figure preparation, and manuscript drafting and critical revision.
\end{itemize}

\noindent
\bibliography{sn-bibliography}

\end{document}